\documentclass[a4paper, 11pt]{article}

\pdfoutput=1

\usepackage{graphicx}

\def\b{\begin{eqnarray}}
\def\e{\end{eqnarray}}
\def\n{\noindent}
\begin{document}

\begin{center}

{\huge \textbf{Equation of State for a van der Waals \vskip.3cm  Universe during Reissner--Nordstr\"om
\vskip.3cm Expansion \\}}

\vspace {10mm}
\noindent
{\large \bf Emil M. Prodanov}$^{\ast \, , \, \, \spadesuit}$, \,
{\large \bf Rossen I. Ivanov}$^{\dagger \, , \, \, \spadesuit \, , \, \, \diamondsuit}$,
\vskip.5cm
{\large \bf and V.G. Gueorguiev}$^{\, \ddagger \, , \, \, \clubsuit \, , \, \, \diamondsuit}$
\vskip1cm
\noindent
\begin{tabular}{c}
$\phantom{e}^\spadesuit ${\it \,\, School of Mathematical Sciences, Dublin Institute of Technology, Ireland} \\
\\
$\phantom{e}^\clubsuit ${\it \,\, School of Natural Sciences and Engineering, University of California -- Merced,} \\
{\it Merced CA 95343, USA} \\
\\
$\phantom{e}^\diamondsuit${\it \,\, On Leave of Absence from Institute for Nuclear Research and Nuclear Energy,} \\
$\phantom{e^\diamondsuit}${\it Bulgarian Academy of Sciences, 72 Tzarigradsko Chaussee, Sofia--1784, Bulgaria} \\
\\
$\phantom{.}^\ast$ {\it emil.prodanov@dit.ie} \\
$\phantom{.}^\dagger$ {\it rossen.ivanov@dit.ie} \\
$\phantom{.}^\ddagger$ {\it vesselin@mailaps.org} \\
\end{tabular}
\vskip1cm
 \end{center}

\begin{abstract}
\n
In a previous work [E.M. Prodanov, R.I. Ivanov, and V.G. Gueorguiev, {\it Reissner--Nordstr\"om Expansion},
Astroparticle Physics 27 (150--154) 2007], we proposed a classical model for the expansion of the Universe during the
radiation-dominated epoch based on the gravitational repulsion of the Reissner--Nordstr\"om geometry --- naked
singularity description of particles that "grow" with the drop of the temperature. In this work we model the Universe
during the Reissner--Nordstr\"om expansion as a van der Waals gas and determine the equation of state.

\end{abstract}

\newpage

\section{Introduction}
\n
In 1971, Hawking suggested \cite{hawk} that there may be a very large number of gravitationally collapsed charged objects
of very low masses, formed as a result of fluctuations in the early Universe. A mass of $10^{14}$ kg of these objects
could be accumulated at the centre of a star like the Sun. Hawking treats these objects classically and his arguments for
doing so are as follows \cite{hawk}: gravitational collapse is a {\it classical} process and microscopic black holes can
form when their Schwarzschild radius is greater than the Planck length $(Gh/c^3)^{-1/2} \sim 10^{-35}$ m (at Planck
lengths quantum gravitational effects do not permit purely classical treatment). This allows the existence of collapsed
objects of masses from $10^{-8}$ kg and above and charges up to $\pm 30$ electron units \cite{hawk}. Additionally, a
sufficient concentration of electromagnetic radiation causes a gravitational collapse --- even though the Schwarzschild
radius of the formed black hole is smaller than the photon's Compton wavelength which is infinite.  Therefore, when
elementary particles collapse to form a black hole, it is not the {\it rest} Compton wavelength $hc/mc^2$ that is to be
considered --- one should instead consider the {\it modified} Compton wavelength $hc/E$, where $E \sim kT >\!\!> mc^2$
is the typical energy of an ultra-relativistic particle that went to form the black hole \cite{hawk}. Microscopic black
holes with Schwarzschild radius greater than the modified Compton wavelength $hc/E$,  can form classically and
independently on competing quantum processes. \\
Hawking suggests that these charged collapsed objects may have velocities in the range 50 -- 10 $\!\!$000 km/s and would
behave in many respects like ordinary atomic nuclei \cite{hawk}. When these objects travel through matter, they induce
ionization and excitation and would produce bubble chamber tracks similar to those of atomic nuclei with the same charge.
The charged collapsed objects survive annihilation and, at low velocities (less than few thousand km/s), they may form
electronic or protonic atoms  \cite{hawk}: the positively charged collapsed objects would capture electrons and thus mimic
super-heavy isotopes of known chemical elements, while negatively charged collapsed objects would capture protons and
disguise themselves as the missing zeroth entry in the Mendeleev table. \\
Such ultra-heavy charged massive particles (CHAMPS) were also studied by de Rujula, Glashow and Sarid \cite{glashow}
and considered as dark matter candidates. \\
Dark Electric Matter Objects (DAEMONS) of masses just above $10^{-8}$ kg and charges of around $\pm 10$ electron units
have been studied in the Ioffe Institute and positive results in their detection have been reported \cite{drob} ---
observations of scintillations in ZnS(Ag) which are excited by electrons and nucleons ejected as the relic elementary
Planckian daemon captures a nucleus of Zn (or S). \\
The DAMA (DArk MAtter) collaboration also report positive results \cite{dama} in the detection of such particles using
100 kg of highly radiopure NaI(Tl) detector. \\
Such heavy charged particles can serve as driving force for the expansion of the Universe during the radiation-dominated
epoch in a classical particle-scale model, which we recently proposed \cite{pig}. Along with this type of  particles,
within our model, magnetic monopoles can also play the same role for the expansion of the Universe: it has been
suggested \cite{preskill} that ultra-heavy magnetic monopoles were created so copiously in the early Universe that they
outweighed everything else in the Universe by a factor of $10^{12}$. \\
Our {\it particle-scale} model gives the expected prediction for the behaviour of the scale factor of the
radiation-dominated expanding Universe, $a(\tau) \sim \sqrt{\tau} \,, $ and can be considered as a complement to the
{\it large-scale} Friedmann--Lema\^itre--Robertson--Walker (FLRW) model (see, for example, \cite{rw, mtw}) which describes
the Universe as isotropic and homogeneous, with very smoothly distributed energy-momentum sources modeled as a perfect
fluid, applicable on scales much larger than galactic ones. \\
This recently proposed \cite{pig} classical mechanism for the cosmic expansion models the Universe as a two-component gas. One of the fractions is that of ultra-relativistic "normal" particles of typical
mass $m$ and charge $q$ with equation of state of an ideal quantum gas of massless particles. The other component is
"unusual" --- these are the particles of ultra-high masses $M$ (of around $10^{-8}$ kg and above) and charges $Q$ (of
around $\pm 10$ electron charges and above) --- exactly as those described earlier. \\
For an elementary particle such as the electron, the charge-to-mass ratio is $q/m \sim 10^{21}$ (in geometrized units
$c = 1 = G$), while for the "unusual" particles, $M \, \lower1pt\hbox{\tiny $\stackrel{<}{\sim}$}\, Q$. In view of this,
the general-relativistic treatment of elementary particles or charged collapsed objects of very low masses also
necessitates consideration from Reissner--Nordstr\"om (or Kerr--Newman) viewpoint --- for as long as their charge-to-mass
ratio remains above unity. We also treat the "unusual" particles classically (in line with Hawking's arguments outlined
earlier). That is, the "unusual" particles are modelled as Reissner--Nordstr\"om naked singularities and the expansion
mechanism is based on their gravito-electric repulsion. Instead of the Schwarzschild radius, the {\it characteristic
length} that is to be considered now and compared to the {\it modified} Compton length \cite{hawk}, will be the radius of
the van der Waals-like impenetrable sphere that surrounds a naked singularity (see \cite{cohen} for a very thorough
analysis of the radial motion of test particles in a Reissner--Nordstr\"om field). As shown in \cite{pig}, for
temperatures below $10^{31}\! $ K, the radius of the impenetrable sphere of an "unusual" particle of mass $10^{-8}$ kg and
charge $\pm 10$ electron units is greater than the {\it modified} Compton wavelength of the "unusual" particle itself. \\
Naked singularities have been subject of significant scrutiny for decades. In the 1950s, the Reissner--Weyl repulsive
solution served as an effective model for the electron. Very recently, a general-relativistic model for the classical
electron --- a point charge with finite electromagnetic self-energy, described as Reissner--Nordstr\"om (spin 0) or
Kerr--Newman (spin 1/2) solution of the Einstein--Maxwell equations, --- has been studied by Blinder \cite{blinder}.
Naked singularities are disliked --- hence the Cosmic Censorship Conjecture \cite{penrose} --- but not ruled out --- there
is no mathematical proof whatsoever of the Cosmic Censorship. At least one naked singularity is agreed to have existed
--- the Big Bang --- the Universe itself. Of particular importance in the study of naked singularities are the work of
Choptuik \cite{choptuik}, where numerical analysis of Einstein--Klein--Gordon solutions shows the circumstances under
which naked singularities are produced, and the work of Christodoulo \cite{chr} who proved that there exist choices of
asymptotically flat initial data which evolve to solutions with a naked singularity. The possibility of observing naked
singularities at the LHC has been studied in \cite{casadio} --- for example, a proton-proton collision could result in a
naked singularity and a set of particles with vanishing total charge or with one net positive charge --- an event probably
undistinguishable from ordinary particle production. In a cosmological setting, naked singularities have been well studied
and classified --- see, for example, \cite{ellis}.

\section{Reissner--Nordstr\"om Expansion}
Consider a "normal" particle of specific charge $q/m$, and an "unusual" particle of charge $Q$ such that
$\mbox{sign}(Q) q/m  \ge - 1$. If the "normal" particle approaches the "unusual" particle from infinity, the field of the
naked singularity is characterized by three regions \cite{pig, cohen}:
\vskip.6cm
\begin{center}
\includegraphics[width=7cm]{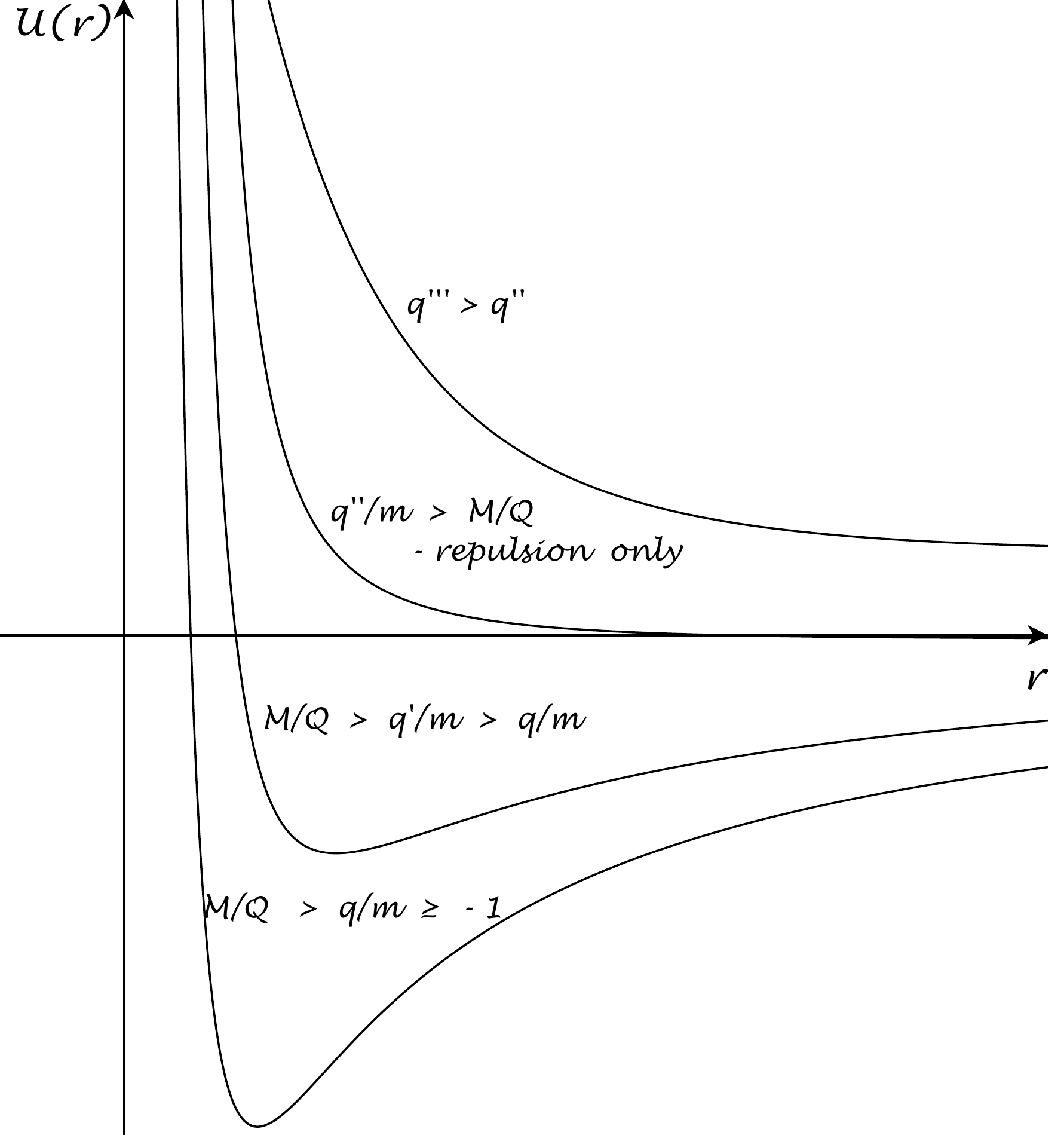}
\vskip.6cm
\parbox{115mm}{\footnotesize  Figure 1: The potential of the interaction between an "unusual" particle of mass $M$ and
charge $Q>0$ and a test particle of specific charge $q/m \ge -1$. If the specific charge of the probe is smaller than
$-1$, then, as shown in \cite{cohen}, the probe will reach the singularity. Note that electrically neutral probes, in
addition to the attraction, also suffer repulsion, while probes of specific charge $q/m > M/Q$ are always repelled
(the gravitational attraction cannot overcome the electric repulsion). The form of this potential is derived later
(\ref{pot}) in this section. The vertical asymptote to each graph is at $r = r_0(T)$ --- the radius of the van der Waals-like
impenetrable sphere surrounding the naked singularity. When minima are present, they are located at the critical radius
$r=r_c$ --- where attraction and repulsion interchange --- see equation (\ref{rc}).}
\end{center}
\begin{itemize}
\item[(a)] {\it Impenetrable region} --- between $r = 0 \, $ and  $ \, r = r_0(T)$. \\
For an incoming test particle, the condition for reality of the kinetic energy leads to the existence of two turning radii
\cite{pig, cohen} with a {\it forbidden} region in-between. The upper (outside) radius, which we denote $r_0(T)$, can be
thought of as a radius of an "impenetrable" sphere surrounding the naked singularity. It depends on the energy of the
incoming particle (or the temperature $T$ of the "normal" fraction of the Universe): the higher the energy (or the
temperature), the deeper the incoming particle will penetrate into the gravitationally repulsive field of the naked
singularity.
\item[(b)] {\it Repulsive region} --- between the turning radius $r_0(T)$ and the critical radius
$r_c \ge r_0(T)$. \\
The critical radius $r_c$ is where the repulsion and attraction interchange (we determine $r_c$ later in this section).
As the temperature drops, the "unusual" particles "grow" (incoming particles have lower and lower energies and turn back
farther and farther from the naked singularity). When the temperature gets sufficiently low, the radius of the "unusual"
particles $r_0(T)$ grows to $r_c$ (but not beyond $r_c$, as the region $r > r_c$ is characterized by attraction and an
incoming particle cannot turn back while attracted). This means that incoming particles have such low energies that they
turn back immediately after they encounter the gravitational repulsion. Incoming particles of charge $q$ such that
$qQ > Mm$ do not even experience attraction --- we shall see that the repulsive region for such particles extends to
infinity (the gravitational attraction will not be sufficiently strong to overcome the electrical repulsion).
\item[(c)] {\it Attractive region} --- from the critical radius $r_c$ to infinity. Again, there is no
gravitationally attractive region for an incoming particle such that $qQ > Mm$.
\end{itemize}
As shown in \cite{cohen}, when an incoming particle has sufficiently large charge which is also opposite in sign to that
of the naked singularity:  $\, \mbox{sign}(Q) q/m  < - 1 \, ,$ the particle will collide with the naked singularity. When
the naked singularity "captures" such particle, its charge $Q$ decreases and its mass $M$ increases. If sufficient number
of incoming particles are captured, $Q$ will eventually become equal to $M$ --- the naked singularity will pick a horizon
and turn into a black hole. This black hole will evaporate quickly afterwards. We will assume that our "unusual" particles
have survived such annihilation. We will also assume that these super-heavy charged particles have survived annihilation
through all other different competing mechanisms --- for example, they could recombine into neutral particles or decay
before or after that (see Ellis et al. \cite{ellis2} on the astrophysical constraints on massive unstable neutral relic
particles and Gondolo et al. \cite{gondolo} on the constraints of the relic abundance of a dark matter candidate --- a
generic particle of mass in the range of $1 - 10^{14}$ TeV, lifetime greater than $10^{14} - 10^{18}$ years, decaying
into neutrinos). \\
An interesting general-relativistic effect (with no classical analogue) is related to the ability of naked singularities
to capture probes of charge having the same sign. This is associated with the inner turning radius which we denote by
$\rho_0(T)$. On Figure 2, the curves representing the two turning radii, $r_0(T)$ and $\rho_0(T)$, are given as functions
of the specific charge $q/m$ of the probe for different temperatures. The forbidden region is between the two curves.
As can be seen, the lower curve $\rho_0(T)$ corresponds to a turning radius (capturing) of a radially outgoing probe
with charge having the same sign as the centre. This has no classical analogue and we argue that it could serve as a
possible mechanism for the formation of the "unusual" particles in the extremely dense very early Universe. Moreover,
this can also allow the extension of the range of validity of our model to account for the inflation of the Universe:
if charge non-conservation of the naked singularities occurs (naked singularities picking up charge), then accelerated
expansion can be achieved: $a(\tau) \sim e^{H\tau}$ or $a(\tau) \sim \tau^n$, with $n > 1$.
\vskip.6cm
\begin{center}
\includegraphics[width=7cm]{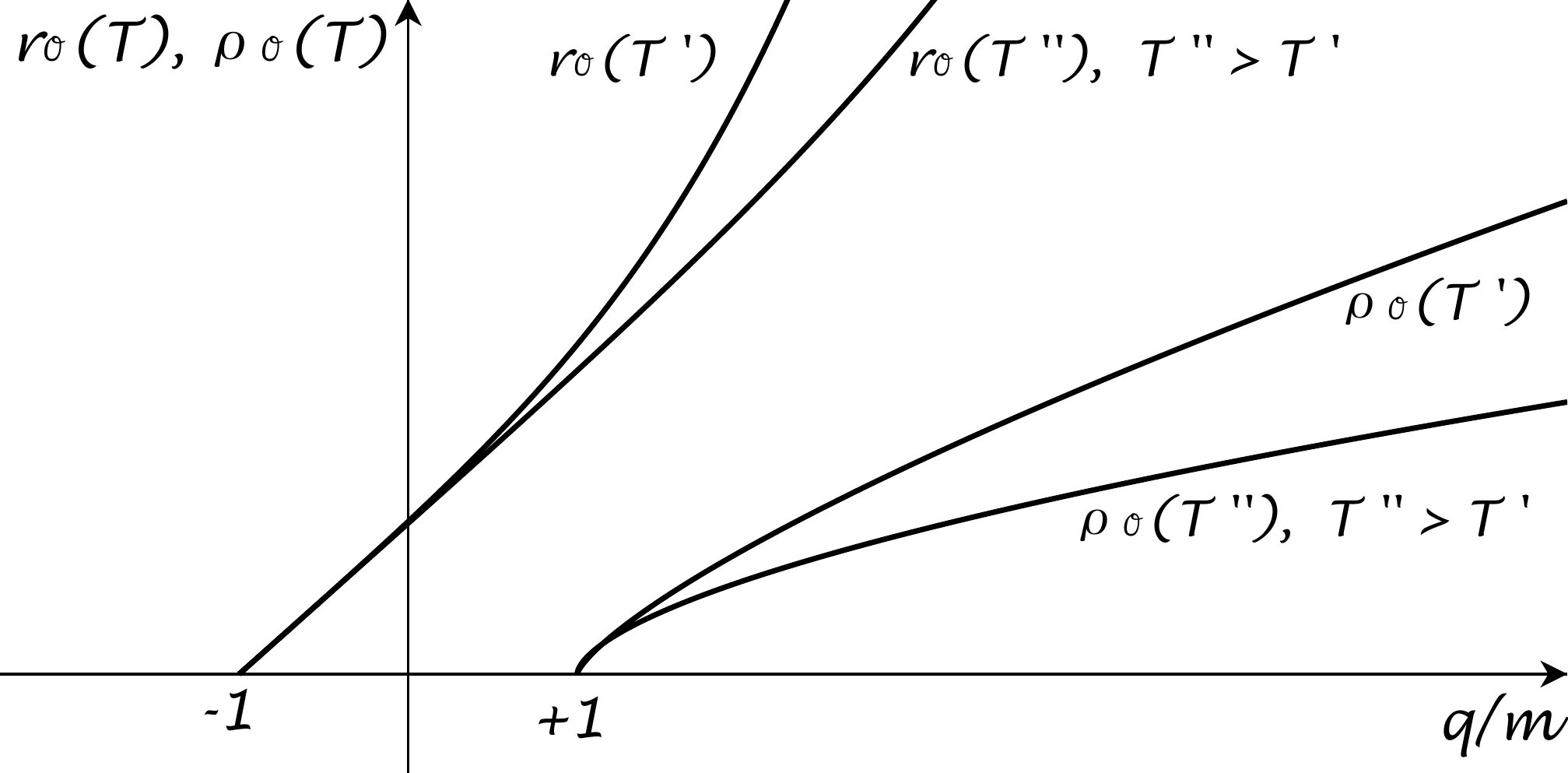}
\vskip.6cm
\parbox{115mm}{\footnotesize  Figure 2: Plotted as functions of the specific charge $q/m$ of a radially moving test
particle in the field of a positively charged naked singularity, the two temperature-dependent curves $r_0(T)$ and
$\rho_0(T)$ represent the outer and inner turning radii, respectively. The region between the curves $r_0(T)$ and
$\rho_0(T)$, for any given temperature $T$, is not allowed as it is characterized by negative kinetic energy. The
explicit form of these curves is given in (\ref{r0}) later in this section (see also \cite{pig, cohen}).}
\end{center}
Our expansion model assumes that initially, at extremely high energies and pressures of the very early Universe, the
"normal" particles are within the gravitationally repulsive regions of the "unusual" particles with radial coordinates
just above the upper turning radius $r_0(T)$. The particles from the "normal" fraction "roll down" the gravitationally
repulsive potentials of the "unusual" particles and in result the Universe expands. The addition of a new class of
particles (the "unusual") in the picture of the Universe does not challenge our current understanding of the physical laws
governing the Universe. The "unusual" particles interact purely classically with the "normal" component of the Universe
and this classical interaction results in the appearance of a repulsive force. Our aim is to offer a possible explanation
for the expansion of the Universe while conforming with the well established theoretical models. As shown in \cite{pig},
during the Reissner--Nordstr\" om expansion, the standard relation between the scale factor of the Universe $a$ and the
temperature $T$ holds: $aT = \mbox{const}$. Also, during the Reissner--Nordstr\" om expansion, the time-dependance of the
scale factor is: $a(\tau) \sim \sqrt{\tau}$ (see \cite{pig} for details). Such is the behaviour of the scale factor during
the expansion of the Universe throughout the radiation-dominated era, obtained by the standard cosmological treatment. \\
On a large scale, the Universe is isotropic and homogeneous and for a FLRW Universe (\cite{rw, mtw}), the
energy-momentum sources are modeled as a perfect fluid, specified by an energy density and isotropic
pressure in its rest frame. This applies for matter known observationally to be very smoothly distributed. On smaller
scales, such as stars or even galaxies, this is a poor description. In our picture, the Universe has global
FLRW geometry, but locally it has Reissner--Nordstr\" om geometry. The compatibility of local
Reissner--Nordstr\" om geometry with global FLRW geometry has been well established: in 1933, McVittie
\cite{mcv} proposed a metric embedding a Schwarzschild solution \cite{schw} in a FLRW universe. In 1993,
Kastor and Traschen (KT) \cite{kt} found a solution desribing a system of an arbitrary number of charged black holes in the
background of a de Sitter universe \cite{ds}. The case of vanishing cosmological constant in the KT solution corresponds
to the static Majumdar--Papapetrou (MP) solution \cite{mp}, while the solution with positive cosmological constant is
highly dynamical and describes black holes exchanging radiation with the background until becoming extreme
($\vert Q \vert = M$). A spinning version of the MP solution with naked singularities was found in \cite{perj} and
\cite{iw}. In 1999, the KT solution was extended \cite{shi} to multi-Kerr-Newman-de Sitter black holes.
Metric for Reissner--Nordstr\"om black holes in an expanding/contracting FLRW universe was obtained in \cite{gao}.
The interplay between cosmological expansion and local attraction in a gravitationally bound system is
studied in \cite{can} where new exact solutions are presented which describe black holes perfectly comoving
with a generic FLRW universe. \\
Returning to the local Reissner--Nordstr\" om geometry, on the level of the interaction between the "unusual" particles and the
"normal" particles of the Universe, the density and pressure variables should be different from those used in the large-scale
 geometry. We are going to complement the entire radiation-dominated era with Reissner--Nordstr\"om
expansion and model the interaction between the "unusual" particles and the "normal" particles as interaction between the
components of a van der Waals gas. Modeling the Universe as a van der Waals phase is possible in the light of the deep
analogies between the physical picture behind the Reissner--Nordstr\" om expansion and the classical van der Waals
molecular model: atoms are surrounded by imaginary hard spheres and the molecular interaction is strongly repulsive in
close proximity, mildly attractive at intermediate range, and negligible at longer distances. The laws of ideal gas must then
be corrected to accommodate for such interaction: the pressure should increase due to the additional repulsion and the
available volume should decrease as atoms are no longer entities with zero own volumes (see, for example, \cite{mandl}). \\
As an interesting development in a similar vein, one should point out the work \cite{cap} (see also the references
therein) which studies van der Waals quintessence by considering a cosmological model comprising of two fluids: baryons,
modelled as dust (large-scale structure fluid) and dark matter with a van der Waals equation of state (background
fluid). Van der Waals equation of state for ultra-relativistic matter has been studied by \cite{bel}. \\
During the Reissner--Nordstr\" om expansion, once the temperature drops sufficiently low so that $r_0(T)$ becomes
equal to $r_c$, the "normal" particle with charge $q$, such that $\mbox{sign}(Q) q/m  \ge - 1 \, $ and also $qQ < mM$,
will be expelled beyond $r = r_c$ (as $r_0(T) < r$ always) --- into the region of gravitational attraction. Due to its
ultra-high energy, the "normal" particle will overcome the gravitational attraction and will escape unopposed to infinity.
Thus the gravitationally attractive region is of no importance for such particles and for them we can assume that the
potential of the naked singularity is infinity from $r = 0$ to $r = r_0(T)$ and zero from $r = r_c$ to infinity. \\
For "normal" particles such that $qQ > mM$, the potential gradually drops to zero towards infinity (there is no attraction
for these probes). For ultra-high temperatures, the energy $E$ of a "normal" particle is of the order of $kT$. At
temperatures below $10^{10}\!$ K, the dominant term in the energy $E$ becomes the particle's rest energy $m c^2$
(throughout the paper we use geometrized units) and, as we shall see, the turning radius $r_0(T)$ becomes infinitely large
below such temperature. As we model the entire radiation-dominated epoch with Reissner--Nordstr\"om repulsion, at
Recombination (the end of this epoch: $t_{recomb} \sim$ 300 $\!$000 years), the free ions and electrons combine to form
neutral atoms ($q = 0$) and this naturally ends the Reissner--Nordstr\"om expansion --- a neutral "normal" particle will
now be too far from an "unusual" particle to feel the gravitational repulsion (the density of the Universe will be
sufficiently low). During the expansion, the volume $V$ of the Universe is proportional to the number $N$ of "unusual"
particles times their volume (one can view the impenetrable spheres of the naked singularities as densely packed spheres
filling the entire Universe). At Recombination, $V \sim t_{recomb}^3 \, $. Therefore, at Recombination, the radius
$r_0(T)$ of an "unusual" particle will be of the order of $R_c = N^{-1/3} t_{recomb} \, $. During the expansion, a "normal"
particle is never farther than $r_0(T)$ from an "unusual" particle. We will request that once $r_0(T)$ becomes equal to
$R_c = N^{-1/3} t_{recomb} \, $, then the potential of the interaction between a naked singularity and a particle of
charge $q$, such that $qQ > mM$, becomes zero. \\
In this paper we use a standard treatment \cite{mandl} to model the van der Waals phase of the Universe as a real gas
and, using the virial expansion, we obtain the gas parameters. Combining the van der Waals equation with
$aT = \mbox{const}$, we find the equation of state describing the classical interaction between the ordinary
particles in the Universe and the ``unusual'' particles. \\
Consider the Reissner--Nordstr\"om geometry \cite{rn, mtw} in Boyer--Lindquist coordinates \cite{bl}:
\b
ds^2 & = & - \, \, \frac{\Delta}{r^2} \, dt^2 + \frac{r^2}{\Delta} \, dr^2  +
r^2 \, d \theta^2 + r^2 \sin^2 \!\theta \, d\phi^2 \, .
\e
where: $\Delta  =  r^2 - 2 M r + Q^2 \, , \, $  $M$ is the mass of the centre, and $Q$ --- the charge of the centre.
We will be interested in the case of a naked singularity only, namely: $Q > M$. \\
The radial motion of a test particle of mass $m$ and charge $q$ in Reissner--Nordstr\"om geometry can be modeled by
an effective one-dimensional motion of a particle in non-relativistic mechanics with the following equation of
motion \cite{pig, cohen} (see also \cite{wald} for Schwarzschild geometry) :
\b
\label{w}
\frac{\dot{r}^2}{2} +  \Bigl[ - \Bigl( 1 - \frac{q}{m} \, \frac{Q}{M} \, \epsilon \Bigr) \frac{M}{r}
+ \frac{1}{2}\Bigl( 1 - \frac{q^2}{m^2} \Bigr) \frac{Q^2}{r^2} \Bigr] = \frac{\epsilon^2 - 1}{2} \, ,
\e
where $\epsilon = E/m$ is the specific energy (energy per unit mass) of the three-dimensional motion. The expression in
the square brackets is the effective non-relativistic one-dimensional potential and the specific energy of the effective
one-dimensional motion is $(1/2)(\epsilon^2 -1)$. As we will not be interested in the effective one-dimensional motion,
we will proceed from equation (\ref{w}) to derive an expression that will serve as gravitational potential energy $U(r)$
of the three-dimensional motion. In the rest frame of the probe ($\dot{r} = 0$), equation (\ref{w}) becomes a quadratic
equation for the energy $\epsilon$. The bigger root of this equation is exactly the gravitational potential energy $U(r)$
plus the rest energy $m$ (see also \cite{def}). Namely:
\b
\label{U}
U(r) \, \, = \, \, \frac{qQ + m \sqrt{\Delta}}{r} - m \, \, = \, \, \frac{qQ}{r} + m \sqrt{1 - \frac{2M}{r} +
\frac{Q^2}{r^2}} - m \, .
\e
Since $M \sim Q \sim 10^{-34}$ cm, expression (\ref{U}) for the potential energy $U(r)$, for distances above
$10^{-34}$ cm, can be approximated by:
\b
\label{pot}
U(r) \, \, = \, \, - \frac{m M}{r} + \frac{q Q}{r} + \frac{m}{2} (-M^2 + Q^2) \frac{1}{r^2} \, .
\e
From now on, we will use this pseudo-Newtonian potential to mimic general-relativistic effects with a classical theory. \\
Motion is allowed only when the kinetic energy is real. Equation (\ref{w}) determines the region $(r_- \, , r_+)$
within which motion is impossible. The turning radii are given by \cite{pig, cohen}:
\b
\label{r0}
r_\pm = \frac{M}{\epsilon^2 - 1} \Biggl[ \epsilon \, \frac{q}{m} \, \frac{Q}{M} - 1
 \pm \sqrt{\Bigl( \epsilon \, \frac{q}{m} \, \frac{Q}{M} - 1 \Bigr)^2
 - (1 - \epsilon^2) \Bigl(1 - \frac{q^2}{m^2} \Bigr) \frac{Q^2}{M^2}} \, \, \Biggr].
 \e
 We identify the impenetrable radius $r_0(T)$ of an ``unusual'' particle as the bigger root $r_+$ and the inner turning
 radius $\rho_0(T)$ as the smaller root $r_-$. The expansion mechanism is based on the fact that $r_0(T)$ is inversely
 proportional to the temperature, namely, the naked singularity drives apart all neutral particles and particles of
 specific charge $q/m$ such that $\mbox{sign}(Q) q/m  \ge - 1$. \\
 Note that when $\epsilon  \to 1$ (which happens when the rest energy becomes the dominant term, i.e. when $kT$ drops below
 $m$, or below $10^{10}$K), then the turning radius $r_0(T)$ tends to infinity. \\
 At the point where gravitational attraction and repulsion interchange, there will be no force acting on the incoming
 particle. That is, this is the point where the derivative of the potential (\ref{pot}) vanishes:
 \b
 \label{rc}
 r_c = M \Bigl(\frac{Q^2}{M^2} - 1 \Bigr)\Bigl(1 - \frac{q}{m} \frac{Q}{M}\Bigr)^{-1}\, .
 \e
 Obviously, the critical radius $r_c$ for an incoming particle charged oppositely to the "unusual" particle ($q Q < 0$)
 will be smaller than the critical radius for a neutral ($q = 0$) incoming particle (neutral particles suffer repulsion)
 as the region of gravitational repulsion will be reduced by the additional electrical attraction. When the incoming probe
 has charge with the same sign as that of the "unusual" particle and $q Q > m M$, then $r_c$ does not exist. This means
 that there will be a region of repulsion only --- the gravitational attraction will not be sufficiently strong to overcome
 the electrical repulsion. \\
 Finally, the potential energy of a charged probe in the field of an ``unusual'' particle can be written as follows:
 \b
 \label{potential}
 U(r) = \left\{
 \begin{array}{ll}
 \infty \, , & \mbox{$r < r_0(T) \, ,$} \cr \cr
 - \frac{mM}{r} + \frac{qQ}{r} + \frac{m}{2} (-M^2 + Q^2) \frac{1}{r^2} \, , &  \mbox{$r_0(T) \le r \le R$ \, ,}
 \cr \cr 0 \, , & \mbox{$r > R \, ,$}
 \end{array}
 \right.
 \e
 where:
 \b
 R = \left\{
 \begin{array}{ll}
 r_c \, , \quad & \mbox{$ \mbox{sign}(Q) q/m  \ge - 1 \, $ and $\, q Q \le m M \, ,$} \cr
 R_{c} \, , \quad & \mbox{$qQ > mM$.}
 \end{array}
 \right.
 \e
 Obviously, the expansion beyond $r_c$ will be due to those particles that satisfy $qQ > mM$.

 \section{Van der Waals Equation of State}
 Next, we consider the thermodynamics of a real gas. The virial expansion relates the pressure $p$ to the particle number
 $N$, the temperature $T$ and the volume $V$
 \cite{mandl}:
 \b
 p = \frac{NkT}{V} \Bigl[ 1 + \frac{N}{V} F(T) + \Bigl( \frac{N}{V} \Bigr)^2 G(T) + \cdots \Bigr] \, ,
 \e
 where the correction term $F(T)$ is due to two-particle interactions, the correction term $G(T)$ is due to three-particle
 interactions and so forth. We will ignore all interactions involving more than two particles. The correction term $F(T)$
 is \cite{mandl}:
 \b
 F(T) = 2 \pi \int\limits_{0}^{\infty} \lambda(r) \, r^2 dr = \beta - \frac{\alpha}{kT} \, ,
 \e
 where $\lambda(r)$ is given by:
 \b
 \lambda(r) = 1 - e^{-\frac{U(r)}{kT}} \, .
 \e
 Then ``van der Waals'' equations is \cite{mandl}:
 \b
 \label{pvdw}
 p + \Bigl( \frac{N}{V} \Bigr)^2 \alpha  =  \frac{NkT}{V} \Bigl( 1 + \frac{N}{V} \beta \Bigr) \, .
 \e
 In the limit $N \beta / V \to 0$, this equation reduces to the usual van der Waals equation \cite{mandl}:
 \b
 \label{vdw}
 \Bigl[ p + \Bigl( \frac{N}{V} \Bigr)^2 \alpha \Bigr]  \Bigl( 1 - \frac{N}{V} \beta \Bigr) =  \frac{NkT}{V}  \, .
 \e
 We now assume that the ``unusual'' particles leave ``voids'' in the Universe where ``normal'' particles cannot enter.
 Thus, the effective space left for the motion of the ``normal'' component of the gas is reduced by $N \beta$, where
 $\beta$ is the ``volume'' of an ``unusual'' particle and $N$ is the number of ``unusual'' particles. We will also
 pretend that ``unusual'' particles are not present and that the potential in which the ``normal'' particles move is not
 due to the ``unusual'' particles, but rather to the two-particle interactions between the ``normal'' component of the
 gas. In essence, we ``remove'' $N$ ``unusual'' particles out of all particles and we are dealing
 with a gas of $n$ ``normal'' particles. The "van der Waals" equation (\ref{pvdw}) then becomes:
 \b
 \label{here}
 p + \Bigl( \frac{N}{V} \Bigr)^2 \alpha = \frac{nkT}{V} (1 + \frac{N}{V} \beta)    \, ,
 \e
 For the potential determined in (\ref{potential}), we have:
 \b
 \lambda(r) = 1 - e^{-\frac{U(r)}{kT}} =
 \left\{
 \begin{array}{ll}
 1 \, , & \mbox{$r < r_0(T) \, ,$} \cr \cr
 \frac{U(r)}{kT} \, , &  \mbox{$r_0(T) \le r \le R \, ,$} \cr \cr
 0 \, , & \mbox{$r > R \, .$}
 \end{array}
 \right.
 \e
 We then get:
 \b
 \label{b}
 \beta & = &  2 \pi \int\limits_{0}^{r_0(T)} \!\! r^2 \, dr
 \, \, = \, \, \frac{2 \pi}{3} r_0^3(T) \, \, = \, \, \frac{1}{2} \, v_0(T) \, , \\
 \label{a}
 \alpha & = & 2 \pi \int\limits_{r_0(T)}^{R} \!\! U(r) \, r^2 \, dr \, \, = \, \,
 \pi m M^2 \Bigl( 1 - \frac{Q^2}{M^2} \Bigr) [R - r_0(T)] \nonumber \\
 & & \hskip100pt + \, \, \pi m M \Bigl( 1 - \frac{q}{m}\frac{Q}{M} \Bigr) [R^2 - r_0^2(T)] \, ,
 \e
 where $v_0(T)$ is the ``volume'' of an "unusual" particle. Note that both $\alpha$ and $\beta$ depend on the temperature
 via the particle's radius $r_0(T)$. \\
 We have shown \cite{pig} that for our expansion model, the standard relation between the scale factor of the Universe
 $a$ and the temperature $T$ holds: $aT = \mbox{const.}$ Let $\rho$ denote the density of the Universe. Then, as the
 volume $V$ of the Universe is proportional to the third power of $a$ and as $V \sim 1/ \rho \, , $ we have
 $T \sim \rho^{1/3}$. Therefore, $T/V \sim \rho^{4/3}$. \\
 The volume $V$ of the Universe during the van der Waals phase is proportional to the volume $v_0(T)$ of the "unusual"
 particles times their number $N$. Using equation (\ref{b}), namely: $\beta = \frac{1}{2} v_0(T)$, it immediately follows
 that $ \, N \beta / V $ is, essentially, constant. \\
 Equation (\ref{here}) is the equation of state for the van der Waals phase of the expanding Universe and can be written:
 as:
 \b
 p = \eta \rho^{4/3} - \frac{\alpha}{\beta^2} \, .
 \e
 Here $\eta$ is some constant. The second term depends on the temperature via $\alpha$ and $\beta$ and becomes irrelevant
 towards the end, as $\alpha \to 0$ when $r_0(T) \to R$. Note also that the correction term $- \alpha / \beta^2$ is
 positive as $\alpha$ is negative.

 \section*{Acknowledgements}
 \noindent
 R.I. Ivanov acknowledges partial support from the Bulgarian National Foundation for Scientific
Investigations, grant 1410.

 \pagebreak

 \end{document}